\documentclass{emulateapj}
\slugcomment{} 
\shorttitle{SDSS QUASAR LENS SEARCH. III.}
\shortauthors{OGURI ET AL.}
\begin{document}
\title{The Sloan Digital Sky Survey Quasar Lens Search. III. Constraints
on Dark Energy from the Third Data Release Quasar Lens Catalog}  
%
\author{
Masamune Oguri,\altaffilmark{1,2} 
Naohisa Inada,\altaffilmark{3,4} 
Michael A. Strauss,\altaffilmark{2} 
Christopher S. Kochanek,\altaffilmark{5} 
Gordon T. Richards,\altaffilmark{6}\\
Donald P. Schneider,\altaffilmark{7} 
Robert H. Becker,\altaffilmark{8,9} 
Masataka Fukugita,\altaffilmark{10} 
Michael D. Gregg,\altaffilmark{8,9} 
Patrick B. Hall,\altaffilmark{11}\\
Joseph F. Hennawi,\altaffilmark{12}
David E. Johnston,\altaffilmark{13,14} 
Issha Kayo,\altaffilmark{15} 
Charles R. Keeton,\altaffilmark{16} 
Bartosz Pindor,\altaffilmark{17}\\
Min-Su Shin,\altaffilmark{2} 
Edwin L. Turner,\altaffilmark{2} 
Richard L. White,\altaffilmark{18}
Donald G. York,\altaffilmark{19} 
Scott F. Anderson,\altaffilmark{20}\\
Neta A. Bahcall,\altaffilmark{2}
Robert J. Brunner,\altaffilmark{21} 
Scott Burles,\altaffilmark{22} 
Francisco J. Castander,\altaffilmark{23} 
Kuenley Chiu,\altaffilmark{24} \\
Alejandro Clocchiatti,\altaffilmark{25} 
Daniel Eisenstein,\altaffilmark{26} 
Joshua A. Frieman,\altaffilmark{19,27,28} 
Yozo Kawano,\altaffilmark{15} 
Robert Lupton,\altaffilmark{2}  \\
Tomoki Morokuma,\altaffilmark{29} 
Hans-Walter Rix,\altaffilmark{30} 
Ryan Scranton,\altaffilmark{31} and
Erin Scott Sheldon\altaffilmark{32} 
}

\altaffiltext{1}{Kavli Institute for Particle Astrophysics and
                 Cosmology, Stanford University, 2575 Sand Hill Road,
                 Menlo Park, CA 94025.} 
\altaffiltext{2}{Princeton University Observatory, Peyton Hall,
                 Princeton, NJ 08544.}  
\altaffiltext{3}{Institute of Astronomy, Faculty of Science, University
                 of Tokyo, 2-21-1 Osawa, Mitaka, Tokyo 181-0015, Japan.} 
\altaffiltext{4}{Cosmic Radiation Laboratory, RIKEN (The Physical and Chemical 
                 Research Organization), 2-1 Hirosawa, Wako, Saitama 351-0198, 
                 Japan.} 
\altaffiltext{5}{Department of Astronomy, The Ohio State University, 
                  Columbus, OH 43210.}                 
\altaffiltext{6}{Department of Physics, Drexel University, 3141
                 Chestnut Street,  Philadelphia, PA 19104.}
\altaffiltext{7}{Department of Astronomy and Astrophysics, The
                  Pennsylvania State University, 525 Davey Laboratory, 
                  University Park, PA 16802.}   
\altaffiltext{8}{IGPP-LLNL, L-413, 7000 East Avenue, Livermore, CA 94550.}
\altaffiltext{9}{Department of Physics, University of California at
                 Davis, 1 Shields Avenue, Davis, CA 95616.}  
\altaffiltext{10}{Institute for Cosmic Ray Research, University of
                 Tokyo, 5-1-5 Kashiwa, Kashiwa City, Chiba 277-8582, Japan.} 
\altaffiltext{11}{Department of Physics and Astronomy, York University,
                  4700 Keele Street, Toronto, Ontario, M3J 1P3, Canada}
\altaffiltext{12}{Department of Astronomy, University of California at
                 Berkeley, 601 Campbell Hall, Berkeley, CA 94720-3411.}
\altaffiltext{13}{Jet Propulsion Laboratory, 4800 Oak Grove Drive,  
                 Pasadena CA, 91109}
\altaffiltext{14}{California Institute of Technology, 1200 East  
                 California Blvd, Pasadena, CA 91125}
\altaffiltext{15}{Department of Physics and Astrophysics, Nagoya
                  University, Chikusa-ku, Nagoya 464-8602, Japan.}
\altaffiltext{16}{Department of Physics and Astronomy, Rutgers University, 
                  Piscataway, NJ 08854.}
\altaffiltext{17}{Space Research Centre, University of Leicester,
                 Leicester LE1 7RH, UK.} 
\altaffiltext{18}{Space Telescope Science Institute, 3700 San Martin Drive, 
                 Baltimore, MD 21218.} 
\altaffiltext{19}{Dept. of Astronomy and Astrophysics, and The Enrico
                 Fermi Institute, 5640 So. Ellis Avenue, The
                 University of Chicago, Chicago, IL 60637.}
\altaffiltext{20}{Astronomy Department, Box 351580, University of
                 Washington, Seattle, WA 98195.}
\altaffiltext{21}{Department of Astronomy, University of Illinois, 1002 West 
                  Green Street, Urbana, IL 61801.}
\altaffiltext{22}{Kavli Institute for Astrophysics and Space Research and 
                  Department of Physics, Massachusetts Institute of
                 Technology, Cambridge, MA 02139.}
\altaffiltext{23}{Institut d'Estudis Espacials de Catalunya/CSIC,
                  Gran Capita 2-4, 08034 Barcelona, Spain.}
\altaffiltext{24}{School of Physics, University of Exeter, Stocker Road, 
                  Exeter EX4 4QL, UK.}
\altaffiltext{25}{Departamento de Astronom\'{i}a y Astrof\'{i}sica, Pontificia 
                  Universidad Cat\'{o}lica de Chile, Casilla 306, Santiago 22, 
                  Chile.}
\altaffiltext{26}{Steward Observatory, University of Arizona, 933 North 
                  Cherry Avenue, Tucson, AZ 85721.}
\altaffiltext{27}{Kavli Institute for Cosmological Physics, University
                 of Chicago, Chicago, IL 60637.}
\altaffiltext{28}{Center for Particle Astrophysics, Fermilab, P.O. Box 500, 
                  Batavia, IL 60510.}
\altaffiltext{29}{National Astronomical Observatory, 2-21-1 Osawa, Mitaka, 
                  Tokyo 181-8588, Japan.}
\altaffiltext{30}{Max Planck Institute for Astronomy, Koenigsstuhl 17, 
                  69117 Heidelberg, Germany}
\altaffiltext{31}{University of Pittsburgh, Department of Physics and
                 Astronomy, 3941 O'Hara Street, Pittsburgh, PA 15260.}
\altaffiltext{32}{Center for Cosmology and Particle Physics, 
                  Department of Physics, New York University, 4
                 Washington Place,  New York, NY 10003.}   

\begin{abstract}
We present cosmological results from the statistics of lensed quasars
in the Sloan Digital Sky Survey (SDSS) Quasar Lens Search. By taking
proper account of the selection function, we compute the expected
number of quasars lensed by early-type galaxies and their image
separation distribution assuming a flat universe, which is then
compared with 7 lenses found in the SDSS Data Release 3 to derive
constraints on dark energy under strictly controlled criteria. For a
cosmological constant model ($w=-1$) we obtain
$\Omega_\Lambda=0.74^{+0.11}_{-0.15}({\rm stat.})^{+0.13}_{-0.06}({\rm
 syst.})$. Allowing $w$ to be a free parameter we find $\Omega_{\rm
 M}=0.26^{+0.07}_{-0.06}({\rm stat.})^{+0.03}_{-0.05}({\rm syst.})$
and $w=-1.1\pm0.6({\rm stat.})^{+0.3}_{-0.5}({\rm syst.})$ when
combined with the constraint from the measurement of baryon acoustic
oscillations in the SDSS luminous red galaxy sample. Our results are
in good agreement with earlier lensing constraints obtained using
radio lenses, and provide additional confirmation of the presence of
dark energy consistent with a cosmological constant, derived
independently of type Ia supernovae.  
\end{abstract}


\section{Introduction}\label{sec:intro}

The accelerating expansion of the universe is one of the central
problems in modern cosmology. This acceleration is usually attributed
to the dominant presence of a negative-pressure component that is
often referred to as  dark energy. There are many models that explain
the acceleration, including a classical cosmological constant, decaying
scalar fields, and topological defects \citep[e.g., see][for a
  review]{peebles03}. In addition it might also be explained by
long-range modifications of the gravitational force law
\citep[e.g.,][]{carroll04}.  

The dark energy is characterized by its cosmological density
$\Omega_{\rm DE}$, and its equation of state $w$, which is defined as 
the pressure divided by the density of dark energy. In particular,
measuring $w$ is a useful test of models for dark energy. 
Since $w$ determines the expansion rate of the universe and the
cosmological distance to a given redshift, not only $\Omega_{\rm DE}$
but also the value of $w$ and its time dependence can be inferred from
distance (or volume) measurements on cosmological scales. One powerful
way to constrain $w$ is via observations of distant type-Ia supernovae
that make use of luminosity-decline rate correlations to standardize
their luminosities \citep{riess98,perlmutter99}: Since the
standardized luminosities of type-Ia supernovae have small 
scatter, they serve as an excellent standard candle to measure
cosmological distances. Another probe of dark energy is
the fluctuation spectrum of the cosmic microwave background
\citep[CMB;][]{debernadis00,spergel03,spergel07}. The integrated
Sachs-Wolfe effect, which can be detected by correlating the 
CMB map with the large-scale distribution of galaxies, allows a
direct detection of the dark energy component
\citep[e.g.,][]{rassat07}. Additional constraints on dark energy come
from baryon acoustic oscillations in the galaxy power spectrum
\citep{eisenstein05,cole05,percival07} and X-ray clusters of galaxies
\citep{allen04,allen07,rapetti05}.  Since different methods involve 
different systematics and degeneracies with the cosmological
parameters, it is of great importance to use as many independent
observations as possible in studying dark energy. 

The statistics of strong lensing offer an alternate constraint on dark
energy \citep[][but see also \citealt{keeton02}]{turner90,fukugita90}. 
The probability that a distant object is strongly lensed is
proportional to the number of possible lensing objects along the line
of sight, and thus quite sensitive to dark energy. This method has
been applied to both optical and radio lens samples to derive
interesting constraints on the value of the cosmological constant
\citep{maoz93,kochanek96,falco98,chiba99}, but such applications have
been limited by the small size of existing lens samples as well as
poor knowledge of source and lens populations \citep{maoz05}. For
instance, past work tended to rule out large values ($\gtrsim 0.7$) of
$\Omega_{\rm DE}$ \citep[e.g.,][]{kochanek96}, because of
overestimates of the luminosity function of galaxies
\citep[e.g.,][]{chiba99}. The most recent lensed quasar survey in the
radio band, the Cosmic Lens All-Sky Survey
\citep[CLASS;][]{myers03,browne03}, contains a statistical sample of
13 lenses. Cosmological constraints from this lens sample are roughly
consistent with the current standard model in which the universe is
dominated by dark energy \citep{chae02,chae03a,chae07,mitchell05}.  

The Sloan Digital Sky Survey Quasar Lens Search
\citep[SQLS;][]{oguri06b} provides a large statistical lens sample
appropriate for studying cosmology. It is based on the optical quasar
catalog from the Sloan Digital Sky Survey \citep[SDSS;][]{york00}, and
therefore is complementary to the CLASS in several ways. In particular
the well-known redshift distribution of quasars and the lensing
selection function \citep[][hereafter Paper I]{oguri06b} allow an
accurate estimate of lensing rates. We present our first complete lens
sample from Data Release 3 \citep[DR3;][]{abazajian05,schneider05} in
\citet[][hereafter  Paper II]{inada07}: It consists of 11 lensed
quasars with flux ratios of faint to bright images greater than
$10^{-0.5}$ (for double lenses) and image separations between $1''$
and $20''$, selected from 22,683 low-redshift ($0.6<z<2.2$) quasars
brighter than $i=19.1$. In this paper, we use a subsample of this
optical lens sample to constrain cosmological parameters, in
particular the dark energy abundance and equation of state.  

This paper is organized as follows. In \S \ref{sec:lens} we
briefly summarize our statistical lens sample. Section \ref{sec:theory}
describes how the expected number of lensed quasars is
computed. We present our results in \S \ref{sec:result}, and summarize
in \S \ref{sec:summary}. We denote the present matter
density as $\Omega_M$. The present dark energy density is described as
$\Omega_{\rm DE}$, or $\Omega_\Lambda$ if the cosmological constant
$w=-1$ is assumed. We use the Hubble constant in dimensionless form
$h=H_0/(100{\rm km\, s^{-1}Mpc^{-1}})$. Throughout the paper we assume
a flat universe $\Omega_M+\Omega_{\rm DE}=1$. Magnitudes quoted in the
paper are corrected for Galactic extinction \citep{schlegel98}.

\section{Lensed Quasar Sample}\label{sec:lens}

Our lensed quasar sample is constructed from the SDSS DR3 spectroscopic
quasar catalog \citep{schneider05}. The properties of the SDSS are
presented in a series of technical papers. \citet{gunn06} describes
the dedicated wide-field 2.5-m telescope. Details of the
photometric survey are given in \citet{fukugita96}, \citet{gunn98},
\citet{lupton99}, \citet{hogg01}, \citet{lupton01}, \citet{smith02},
\citet{pier03}, \citet{ivezic04}, \citet{tucker06}, and \citet{lupton07}.
\citet{blanton03} present the tiling algorithm of the spectroscopic
survey. Spectroscopic quasar targets are selected according to an
algorithm described in \citet{richards02}. Details of each public data
set are given in a series of data release papers
\citep{stoughton02,abazajian03,abazajian04,abazajian05,adelman06,adelman07}. 

\begin{deluxetable*}{ccccccccc}
\tablecaption{SDSS DR3 Quasar Lens sample\label{tab:lens}}
\tablewidth{0pt}
\tablehead{
 \colhead{Name}  & \colhead{$N_{\rm img}$} &\colhead{$z_s$} &
 \colhead{$z_l$} & \colhead{$i_{\rm cor}$} & \colhead{$\theta$} &
 \colhead{Lens}  & \colhead{Note} & \colhead{Ref.}}
\startdata
SDSS J0246$-$0825& 2 & 1.685 & 0.723  & 17.77 & 1.04 & E & & 1, 2\\
SBS0909+523      & 2 & 1.377 & 0.83   & 16.17 & 1.11 & E & & 3, 4\\
SDSS J0924+0219  & 4 & 1.523 & 0.393  & 18.12 & 1.78 & E & & 5, 6, 7\\
SDSS J1001+5027  & 2 & 1.839 & \nodata& 17.31 & 2.86 & E?& & 8\\
SDSS J1021+4913  & 2 & 1.720 & \nodata& 18.97 & 1.14 & ? & & 9\\
SDSS J1226$-$0006& 2 & 1.125 & 0.517  & 18.23 & 1.24 & E & & 10, 11\\
SDSS J1335+0118  & 2 & 1.571 & 0.440  & 17.53 & 1.57 & E & & 12, 11\\ \hline
Q0957+561        & 2 & 1.413 & 0.36   & 16.67 & 6.17 & E &
$\theta>3''$ & 13, 14\\
SDSS J1004+4112  & 5 & 1.740 & 0.68   & 18.84 & 14.6 & C &
$\theta>3''$ & 15, 16\\
SDSS J1332+0347  & 2 & 1.438 & 0.191  & 17.89 & 1.14 & E & $i_{\rm
  qso}-i_{\rm gal}>0$ & 17\\
SDSS J1524+4409  & 2 & 1.210 & 0.310  & 18.76 & 1.67 & E & $i_{\rm
  qso}-i_{\rm gal}>0$ & 18\\
\enddata
\tablecomments{See Paper II for details of the construction of the
  statistical lens sample. $i_{\rm cor}$ is the $i$-band PSF magnitude 
  of the object with SDSS spectroscopy, corrected for Galactic
 extinction. The image separation $\theta$ is defined by the maximum
 separation between any image pairs. ``Lens'' indicates the morphology
 of the lensing galaxy: S=spiral; E=elliptical; C=cluster;
 ?=unknown. We adopt the first 7 lenses for our statistical study; the
4 lower lenses are not used because of reasons indicated in the Note.}     
\tablerefs{(1) \citealt{inada05}; (2) \citealt{eigenbrod07};
(3) \citealt{oscoz97}; (4) \citealt{lubin00};
(5) \citealt{inada03a}; (6) \citealt{ofek06}; (7) \citealt{eigenbrod06a};
(8) \citealt{oguri05a}; (9) \citealt{pindor06}; 
(10) \citealt{inada03c}; (11) \citealt{eigenbrod06b};
(12) \citealt{oguri04b}; (13) \citealt{walsh79}; 
(14) \citealt{young81}; (15) \citealt{inada03b};
(16) \citealt{oguri04a}; (17) \citealt{morokuma07};
(18) \citealt{oguri07b}
}
\end{deluxetable*}

The DR3 statistical lens sample contains 11 lensed quasars (see Paper
II). The sample is restricted to a range of $i$-band flux ratios,
which are the fluxes of the fainter images divided by those of the
brighter images for double lenses (no condition on flux ratios is
set for quadruple lenses), $f_i>10^{-0.5}$ and image separations
$1''<\theta<20''$ where the completeness of the candidate selection is
almost unity (see Paper I). In this paper, we apply two additional
cuts to select a subsample appropriate for our dark energy study. 
First, we restrict the image separation range to $1''<\theta<3''$.  At
$\theta<3''$ lens potentials are in many cases dominated by those from
individual lensing galaxies, whereas beyond $\theta=3''$ the
contribution of surrounding dark matter to lens potentials begins to
become more significant \citep[e.g.,][]{kochanek01,oguri05b,oguri06a}. 
The effect of the external field  can, in principle, be included in
our theoretical model of lensing rates. However, it is difficult to
observationally constrain the probability distribution of external
fields (note that we adopt observationally determined velocity
functions for our computations. See \S \ref{sec:gal}), indicating that
it introduces additional systematics. Second, we require that the
lensing galaxy be fainter than the quasar components. If the lens
galaxy is too bright, it will strongly affect the colors of the
quasars and the lensed quasar will not be targeted for spectroscopy
\citep{richards02}, biasing against us discovering the lens. In
addition, lens galaxy fluxes add to the brightness of the system,
which could enhance the number of lenses in the flux-limited
sample. These biases make theoretical predictions much more difficult
and uncertain.  Therefore, we require that the $i$-band point spread
function (PSF) magnitude for the quasar components, $i_{\rm qso}$, be
brighter than the $i$-band magnitude of the lensing galaxy, $i_{\rm
  gal}$. These two cuts remove four lensed quasars from the lens
sample. Hereafter, we use the remaining seven lensed quasars to
constrain cosmological parameters. Table \ref{tab:lens} summarizes the
lens sample we use in this paper.  

\section{Theoretical Model}\label{sec:theory}

In this section, we describe how to compute the expected number of
lensed quasars in the SQLS DR3 sample in a cosmological model,
following \citet{turner84} with the selection function taken into
account.   
 
\subsection{Lens Galaxy Population}\label{sec:gal}
We consider early-type galaxies as lensing objects. Although late-type
galaxies are more abundant, standard models predict that the strong
lensing probability is dominated by that from early-type galaxies
\citep[e.g.,][]{turner84,maoz93,kochanek96,moeller07}. This is
particularly true if we restrict image separations to be larger than
$1''$, because late-type galaxies have smaller velocity dispersions on
average and therefore have smaller mean image separations.  Indeed
this is confirmed by observations. Only a few of the $\gtrsim 60$
known lensed quasars with $\theta>1''$ are produced by late-type
galaxies.\footnote{See the CASTLES webpage at
  http://cfa-www.harvard.edu/castles/}  Moreover, none of the lensed
quasars in our sample appear to be caused by late-type galaxies (see
Table \ref{tab:lens}). 

The contribution of large-scale dark matter fluctuations around
lensing galaxies (environmental convergence) is important because it
could bias cosmological results from lensing statistics
\citep{keeton04}. The maximum image separation of $3''$, however,
makes the effect of external convergence due to associated dark matter
on the lensing probability moderate. Specifically, the environmental
convergence enhances the integrated lensing probability only by
$\lesssim 10\%$ at $\theta<3''$ \citep{oguri05b}. Therefore we 
neglect external convergence, although we examine its impact on our
results below. 

It has been argued that the radial mass profile of galaxies can be 
approximated as a singular isothermal sphere for the scales
relevant for strong lensing \citep[e.g.,][]{rusin05,koopmans06}. In
this paper, we adopt an elliptical version of this, a Singular
Isothermal Ellipsoid (SIE). The ellipticity does not significantly
affect the total lensing cross section \citep{huterer05}, but
including ellipticities allow one to take account of the different
selection functions of double and quadruple lenses (see Paper I). The 
two-dimensional surface mass density of an SIE with ellipticity $e$ at 
a position $x$ and $y$ from the center of the galaxy, with the $x$
axis aligned with the major axis, is given by 
\begin{equation}
\kappa(x,y)=\frac{\theta_{\rm E}\lambda(e)}{2}
\left[\frac{1-e}{(1-e)^2x^2+y^2}\right]^{1/2},
\end{equation}
where $\theta_{\rm E}$ denotes the Einstein radius, which in turn is
related to the galaxy velocity dispersion $\sigma_v$ by 
\begin{equation}
\theta_{\rm E}=4\pi\left(\frac{\sigma_v}{c}\right)^2\frac{D_{\rm ls}}{D_{\rm os}},
\end{equation}
where $D_{\rm ls}$ and $D_{\rm os}$ are the angular diameter distances
from lens to source and from observer to source, respectively. 
The parameter $\lambda(e)$ is the velocity dispersion normalization
factor for non-spherical galaxies. 

It is not straightforward to determine the normalization factor
$\lambda(e)$. What is needed is the calculation of velocity dispersions
for lens galaxies, which depends on three-dimensional shape of the
lensing galaxies when the assumption of spherical symmetry is relaxed.
Two extreme possibilities are that all galaxies have either
oblate or prolate shapes. In this paper, we assume that there are
equal number of oblate and prolate galaxies and compute the
normalization by taking the average of normalizations in the oblate
and prolate cases \cite[see][]{chae03a}. This is a reasonable
assumption given the fact that underlying dark halos have triaxial
shapes \citep{jing02}. Moreover the observed axis ratio distribution
is consistent with a population of triaxial early-type galaxies 
\citep{vincent05}. For the distribution of ellipticities, we adopt
a Gaussian distribution with mean $\bar{e}=0.3$ and dispersion
$\sigma_e=0.16$ (but truncated at $e=0$ and $0.9$) that is consistent
with the observed ellipticity distributions of the light of early-type
galaxies \citep{bender89,saglia93,jorgensen95,rest01,alam02,sheth03}. 

One of the most important elements in predicting the number of lensed
quasars is the velocity function of galaxies. As a fiducial velocity
function we adopt that derived from the latest SDSS DR3
data \citep{choi07}, which is fitted by a modified  Schechter function
of the form 
\begin{equation}
\frac{dn}{d\sigma_v}=\phi_*\left(\frac{\sigma_v}{\sigma_*}\right)^\alpha
\exp\left[
  -\left(\frac{\sigma_v}{\sigma_*}\right)^\beta\right]\frac{\beta}{\Gamma 
(\alpha/\beta)}\frac{d\sigma_v}{\sigma_v},
\end{equation}
where ($\phi_*$, $\sigma_*$, $\alpha$, $\beta$)$=$($8.0\times
10^{-3}h^3{\rm Mpc^{-3}}$, $161\,{\rm km\,s^{-1}}$, $2.32$, $2.67$).
This is somewhat different from the velocity function derived by
\citet[][see also \citealt{mitchell05}]{sheth03} upon SDSS DR1
data. They fit the same functional form, but found different
parameters, ($\phi_*$, $\sigma_*$, $\alpha$, $\beta$)$=$($4.1\times 
10^{-3}h^3{\rm Mpc^{-3}}$, $88.8\,{\rm km\,s^{-1}}$, $6.5$, $1.93$). 
We will use the \citet{sheth03} parameters to estimate the size of the
systematic error due to the velocity function.  

Since the lens galaxy will typically have $D_{\rm ls}\sim 0.5D_{\rm
os}$, the lenses in our survey can have redshifts up to $z\sim 1$
(e.g., SBS0909+523 has the lens redshift of $0.83$), and
thus any redshift evolution of the velocity function, which could
change the lensing rate and degenerate with cosmology
\citep{keeton02}, must be taken into account.  While it has been
argued that early-type galaxies evolve only through passive luminosity
evolution at $z\lesssim 1$ \citep[e.g.,][]{im02,willmer06},
theoretical studies favor slight evolution with redshift through
mergers \citep[e.g.,][]{newman00}.  On the other hand the lens
redshift distribution of strong lensing is consistent with no
evolution of the velocity function
\citep{ofek03,chae03b,mitchell05,capelo07}. In this paper we assume
that the velocity function does not evolve, but we also consider the
evolution model used in \citet[][based on a semi-analytic model by
  \citealt{kang05}]{chae07} as well to estimate the systematic impact
of the evolution on our result. In the model the number density and
the characteristic velocity dispersion are simply replaced by
$\phi_*\rightarrow \phi_*(1+z)^{-0.229}$ and $\sigma_*\rightarrow
\sigma_*(1+z)^{-0.01}$.   

\subsection{Quasar Luminosity Function}\label{sec:qso}
The quasar luminosity function is used to calculate the
magnification bias. We adopt the luminosity function constrained from
the combination of the SDSS and 2dF (2SLAQ), namely the 
2SLAQ+\citet{croom04} model in \cite{richards05}, as our fiducial
model:  
\begin{equation}
\Phi(M_g)=\frac{\Phi_*}{10^{0.4(1-\beta_{\rm h})(M_g-M_g^*)}+
10^{0.4(1-\beta_{\rm l})(M_g-M_g^*)}},
\end{equation}
\begin{equation}
M_g^*(z)=M_g^*(0)-2.5(k_1z+k_2z^2),
\end{equation}
with the parameters of ($\beta_{\rm h}$, $\beta_{\rm l}$, $\Phi_*$, 
$M_g^*(0)$, $k_1$, $k_2$)$=$(3.31, 1.45, $1.83\times10^{-6}(h/0.7)^3{\rm
 Mpc}^{-3}{\rm mag}^{-1}$, $-21.61+5\log(h/0.7)$, 1.39, $-0.29$). The
bright and faint end slopes are broadly consistent with those in the
bolometric luminosity function of \citet{hopkins07} at $z\sim 1-2$.
The luminosity function is in terms of rest-frame $g$-band absolute
 magnitudes: We convert it to observed $i$-band apparent magnitudes
 using the K-correction derived in \citet{richards06}. Since the
 luminosity 
 function was derived assuming $\Omega_M=0.3$ and $\Omega_\Lambda=0.7$,
 we adopt this cosmology for computing the absolute magnitudes used to
 compute the magnification bias no matter what values of $\Omega_{\rm M}$
 and $w$ we consider for the remainder of the analysis. 

\subsection{Number of Lensed Quasars}

The lensing cross section $\sigma_{\rm lens}$ for a given lens is
computed numerically using the public code {\it lensmodel}
\citep{keeton01}. We compute the sum: 
\begin{equation}
\sigma_{{\rm lens},i}=\int d{\bf u} \frac{\Phi(L/\mu)}{\mu\Phi(L)},
\label{eq:mb}
\end{equation}
over the source plane positions ${\bf u}$ (with magnification $\mu$)
where multiple images are produced. The cross sections are computed in
units of $\theta_{\rm E}$ and they are weighted by the ratio of the 
differential luminosity functions in order to take magnification bias
into account. The suffix $i$ indicates the number of images, with 
$i=2$ for double lenses and $i=4$ for quadruple lenses. For double
lenses the integral is performed over the region where the
flux ratio of faint to bright images is larger than $10^{-0.5}$, in
order to match the selection function of our lensed quasar
sample. In Paper I, we found that the magnification factor of lensed
images depends on the image separation, because the SDSS quasar target
selection adopts the PSF magnitude for the magnitude limit. 
It was found that the dependence is fitted well by the following form
(see Paper I for details): 
\begin{equation}
\mu=\bar{\mu}\mu_{\rm tot}+(1-\bar{\mu})\mu_{\rm bri},
\label{eq:mag}
\end{equation}
and
\begin{equation}
\bar{\mu}=\frac{1}{2}\left[1+\tanh\left(1.76-1.78\theta\right)\right],
\end{equation}
where $\theta$ is in units of arcsecond and $\mu_{\rm tot}$ and
$\mu_{\rm bri}$ are the total magnification and magnification of the
brightest image, respectively.\footnote{We note that eq. (15) of Paper
  I holds only approximately if the slope of the source luminosity
  function is close to $-2$. In this paper we compute the full
  magnification bias using eqs. (\ref{eq:mb}) and (\ref{eq:mag}).} We
compute the lensing cross section as a function of dimensionless image 
separation $\hat{\theta}=\theta/\theta_{\rm E}$, i.e., $d\sigma_{{\rm  
lens},i}/d\hat{\theta}$.

From the lensing cross section, we can compute the differential
probability that a source at $z=z_s$ is strongly lensed with the image
separation $\theta$ as 
\begin{eqnarray}
\frac{dp_i}{d\theta}(z_s,i_{\rm qso})&=&
\phi_i(\theta)\int_0^{z_s}dz_l 
 \frac{c\,dt}{dz_l}(1+z_l)^3\int d\sigma_v\frac{dn}{d\sigma_v}\nonumber\\
&&\hspace*{-15mm}\times
 \int d\hat{\theta} (D_{\rm ol}\theta_{\rm E})^2  \frac{d\sigma_{{\rm
      lens},i}}{d\hat{\theta}} 
\delta(\theta_{\rm E}\hat{\theta}-\theta)
\Theta(i_{\rm gal}-i_{\rm qso})
\nonumber\\
&=&\phi_i(\theta)\int_0^{z_s}dz_l 
 \frac{c\,dt}{dz_l}(1+z_l)^3\nonumber\\
 &&\hspace*{-15mm}\times\int \frac{d\hat{\theta}}{\hat{\theta}}\frac{d\sigma_v}{d\theta_{\rm E}}
 \frac{dn}{d\sigma_v}(D_{\rm ol}\theta_{\rm E})^2\frac{d\sigma_{{\rm
 lens},i}}{d\hat{\theta}}\Theta(i_{\rm gal}-i_{\rm qso}), 
\label{eq:dpdt}
\end{eqnarray}
where $\phi_i(\theta)$  is the completeness of our lens candidate
selection estimated from simulations of the SDSS images (see Paper I;
for double lenses we adopt the completeness averaged over the flux ratio
between $10^{-0.5}$ and $1$), $c\,dt/dz_l$ denotes the proper
differential distance at $z_l$, and $(D_{\rm ol}\theta_{\rm E})^2$
converts the lensing cross section from the dimensionless unit to the
physical unit. The Heaviside step function $\Theta(i_{\rm gal}-i_{\rm
qso})$ is added to include the condition that the quasar components
should be brighter than the lensing galaxy. We compute $i_{\rm gal}$
using the correlation between the luminosity and velocity dispersion
of early-type galaxies measured by \citet{bernardi03} combined with
K-correction from \citet{fukugita95}. The effect of the luminosity
evolution measured by \citet{bernardi03} is included. The estimated
galaxy magnitudes are broadly consistent with those expected from the
empirical scaling observed for known lenses \citep{rusin03}. The total
lensing probability for image separations between $1''$ and $3''$ is
then given by  
\begin{equation}
p_i(z_s,i_{\rm qso})=\int_{1''}^{3''}d\theta \frac{dp_i}{d\theta}(z_s,i_{\rm qso}). 
\end{equation}

The expected number of lensed quasars in our quasar sample is computed 
by counting the number of quasars weighted by the lensing
probability. Ultimately this can be done by adding the probabilities for
all source quasars:
\begin{equation}
N_i=\sum_{\rm source \, QSOs}p_i(z_s,i_{\rm qso}).
\end{equation}
To save computational time, we actually calculate the expected number 
of lensed quasars for each redshift-magnitude bin and then sum over
the bins. If $N_{\rm qso}(z_{s,j}, i_{{\rm qso},k})$ is the number
of source quasars in the redshift range $z_{s,j}-\Delta
z_s/2<z_s<z_{s,j}+\Delta z_s/2$  and a magnitude range $i_{{\rm
qso},k}-\Delta i/2<i_{\rm qso}< i_{{\rm qso},k}+\Delta i/2$, 
then the predicted number of lensed quasars is 
\begin{equation}
N_i=\sum_{z_{s,j}}\sum_{i_{{\rm qso},k}}N_{\rm qso}( z_{s,j}, i_{{\rm
    qso},k}) p_i(z_{s,j}, i_{{\rm qso},k}).
\label{eq:numlens}
\end{equation}
We adopt bin widths of $\Delta z_s=0.1$ and $\Delta i=0.2$.

\subsection{Likelihood}
We perform a likelihood analysis on the DR3 SQLS lens
sample. The likelihood is computed using the method
introduced by \citet{kochanek93}  
\begin{eqnarray}
  \ln L &=& \ln \left[\prod_{\rm lens} \frac{dp_i}{d\theta}\prod_{\rm
		 unlensed\,QSOs}(1-p_2-p_4)\right] \nonumber\\
&\simeq& \sum_{\rm lens} \ln\left(\frac{dp_i}{d\theta}\right)-(N_2+N_4),
\label{eq:like}
\end{eqnarray}
where $dp_i/d\theta$ is calculated from eq. (\ref{eq:dpdt}) and $N_2$
(doubles) and $N_4$ (quadruples) are from eq. (\ref{eq:numlens}). 
We note that the valid approximation $p_2$, $p_4\ll 1$ is used here.
We neglect three image events caused by naked cusps because they make a
negligible contribution to the total cross section. The summation in
the first term runs over the seven lensed quasars 
listed in Table \ref{tab:lens}. The distribution of lens redshifts
offers an independent test of cosmological model, however it has been
shown that it is more sensitive to the redshift evolution of the lens
galaxy population as well as the selection bias
\citep[e.g.,][]{ofek03,capelo07}. We do not include information on the
lens redshift $z_l$ (i.e., we adopt the probability after integrating
over the lens redshift in eq. [\ref{eq:dpdt}]) because the lens 
redshifts are not known for all lenses (see Table \ref{tab:lens}) and
including $z_l$ for those objects that do have lens redshifts might
introduce systematic effect related to the incompleteness of our 
redshift information. 

\section{Results}\label{sec:result}

In this section, we derive constraints on cosmological parameters from
the lens statistics. Since the abundance of  lensing galaxies we adopt
in this paper was determined observationally, our results do not depend
on the normalization $\sigma_8$ or shape of the primordial power
spectrum. The Hubble constant is not important for these calculation
because no absolute length scale is used. We assume a flat universe
$\Omega_M+\Omega_{\rm DE}=1$, therefore there are only two independent
cosmological parameters in our analysis, $\Omega_{\rm DE}$
($\Omega_\Lambda$) and $w$.   

\begin{figure}
\epsscale{.99}
\plotone{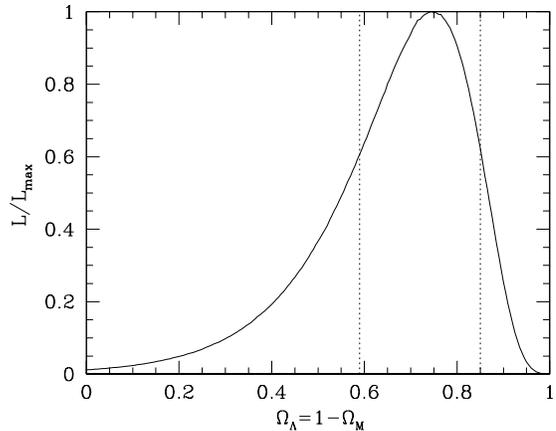}
\caption{Relative likelihoods of the value of the cosmological constant
  $\Omega_\Lambda$ from fitting the SQLS DR3 data, assuming
  a spatially flat universe. The vertical dotted lines indicate
  the $1\sigma$ range estimated from $\Delta\chi^2=1$. The likelihood
  becomes maximum at $\Omega_\Lambda=0.74$.  
\label{fig:lambda}} 
\end{figure}

\subsection{Cosmological constant}\label{sec:cc}

First we derive constraints on the cosmological constant
$\Omega_\Lambda$ assuming $w=-1$. We compute the likelihood
(eq. [\ref{eq:like}]) as a function of $\Omega_\Lambda$, which is 
plotted in Figure \ref{fig:lambda}. The resulting constraint
$\Omega_\Lambda=0.74^{+0.11}_{-0.15}$ ($1\sigma$; the error is estimated
from $\Delta\chi^2\equiv -2\ln (L/L_{\rm max})=1$) is broadly
consistent with other measurements of the cosmological constant
\citep[e.g.,][and references therein]{spergel07}. It is also
consistent with  CLASS, for which the best-fit cosmological constant
was $\Omega_\Lambda=0.7-0.8$ \citep{chae03a,chae07,mitchell05}. 
The case $\Omega_\Lambda=0$ has $\Delta\chi^2\sim 9$ (the expected
total number of lenses in the DR3 sample is $\sim 1$) and is therefore  
rejected at the $3\sigma$ level.

\begin{figure}
\epsscale{.99}
\plotone{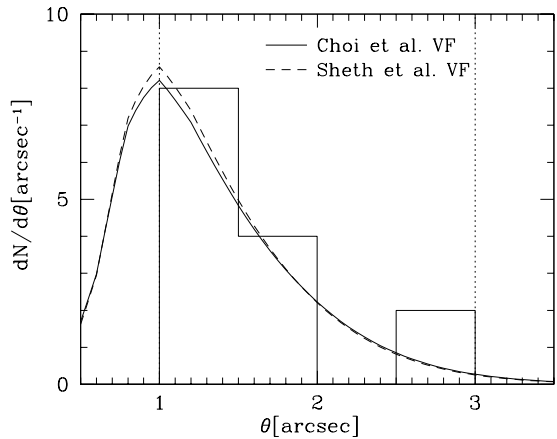}
\caption{The number distribution of lensed quasars is plotted as a
 function of the image separation $\theta$. The histogram shows the
 number distribution in the SQLS DR3 statistical sample (see Table
 \ref{tab:lens}. The bin size is $0\farcs5$, thus the actual
 number of lenses in each bin is half of what we plot). We use only
 lenses in the image separation range $1''< \theta<3''$ as 
 indicated by the vertical dotted lines. The solid line indicates the
 prediction of our best-fit model $\Omega_\Lambda=0.74$ (see Figure
 \ref{fig:lambda}). The dashed line shows the prediction of our
 best-fit model when we adopt the velocity function of \citet{sheth03}
 instead of our fiducial velocity function of \citet{choi07}. See \S
 \ref{sec:syst} for a detailed discussion of the effect of adopting
 the different velocity functions. The sharp decline below
 $\theta=1''$ is due to the selection function $\phi_i(\theta)$, which
 rapidly decreases at $\theta<1''$.  Note that our statistical lens
 sample contains two more lensed quasars at $\theta>3''$ that are not
 shown in this figure. 
\label{fig:sep}} 
\end{figure}

We test the validity of our best-fit model ($\Omega_\Lambda=0.74$) by
comparing the image separation distribution with the observation. In
Figure \ref{fig:sep} we plot the expected number distribution
(eq. [\ref{eq:dpdt}] summed over all sources) with the image
separations of 7 lensed quasars in the statistical sample. Although
the small number of lensed quasars prevents detailed comparison, both
the normalization and the overall shape of the curve appear to match
the observed one. The Kolmogorov-Smirnov test finds that the best-fit
model is consistent with the observed distribution at a significance
level of 72\%.  

\subsection{Dark energy}\label{sec:de}

\begin{figure}
\epsscale{.99}
\plotone{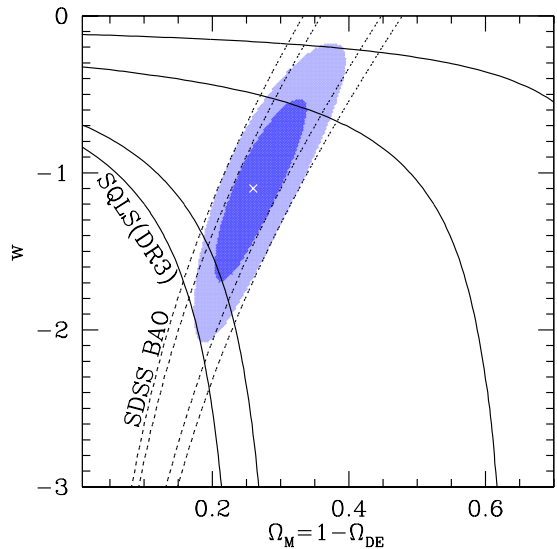}
\caption{Contours at 1$\sigma$ and 2$\sigma$ confidence levels
 (estimated from $\Delta\chi^2=2.3$ and $6.17$) are
 plotted in the $\Omega_{\rm M}$-$w$ plane. Solid lines indicate
 the constraint from the SQLS DR3, whereas dotted lines are from the
 baryon acoustic oscillations (BAO) detected in the SDSS luminous red
 galaxy power spectrum \citep{eisenstein05}. The joint constraint from
 SQLS and BAO are shown by shaded regions: The best-fit model
 ($\Omega_{\rm M}$, $w$)$=$($0.26$, $-1.1$) is indicated with a
 cross. 
\label{fig:cont}} 
\end{figure}

Next we derive constraints on the equation of state $w$ as well as the
dark energy abundance $\Omega_{\rm DE}$. To do so we compute the
likelihood as a function of $\Omega_{\rm M}=1-\Omega_{\rm DE}$ and $w$,
and compute constraints in the two-parameter space. The result is shown
in Figure \ref{fig:cont}. The confidence region shows degeneracy in a
similar direction as constraints from type Ia supernovae
\citep[e.g.,][]{astier06,wood07}, therefore strong lensing alone does
not constrain these parameters very well. As a complementary
constraint, we also consider the likelihood from the measurement of
the scale of baryon acoustic oscillations (BAO) in the SDSS luminous red
galaxy power spectrum \citep{eisenstein05}. Specifically we adopt their
constraint $A\equiv D_V(0.35)\sqrt{\Omega_MH_0^2}/0.35c=0.469\pm0.017$,
where $D_V(0.35)$ is the dilation scale at $z=0.35$. From the combined
constraint shown in Figure \ref{fig:cont}, we obtain $\Omega_{\rm
M}=0.26^{+0.07}_{-0.06}$ and $w=-1.1\pm0.6$ ($1\sigma$). Again the
constraint is consistent with other measurements of dark energy
\citep[e.g.,][]{tegmark06,astier06,spergel07,wood07}. 

\subsection{Systematic errors}\label{sec:syst}

In this paper we have made a number of assumptions to compute the
expected number of lensed quasars in each cosmological model. Here we
examine the sensitivity of our results to these assumptions. To do so,
we change the input model within the expected uncertainties, and compute 
the best-fit cosmological parameters to determine their sensitivity to
these details. We consider the following sources of systematic errors. 

\begin{itemize}
\item For the mass distribution of the lens galaxy, we change the
  fraction of prolate/oblate shapes by $\pm 25\%$ to derive a rough
  estimate of the 
systematic error coming from the dynamical normalization. We also
change the peak of the ellipticity distribution by $\pm0.1$, roughly
corresponding to the current uncertainty in observations
\citep{bender89,saglia93,jorgensen95,rest01,alam02,sheth03}. 
\item  We change the faint end slope of the quasar luminosity function
  by $\pm0.2$ while keeping the other parameters fixed, in order to
  examine the sensitivity to the quasar luminosity function.
 We experiment with the faint end slope because the most poorly
  constrained part of the quasar luminosity function is the faint end
  slope with some evidence that it is shallower than in our model
  \citep{jiang06}. The observed quadruple lens fraction in the SQLS
  also favors slightly shallower faint end slope \citep{oguri07a}.
\item In this paper we have neglected the contribution of external
convergence and shear \citep{oguri05b}: To see how the external fields
change the predicted number of lenses we simply assume the
redshift-independent distributions of external fields are
$0.05\pm0.2$~dex for shear and $0.01\pm0.5$~dex for convergence
\citep[e.g.,][see also \citealt{momcheva06}]{dalal04}. 
\item For the velocity function, we replace the
function we use with that of \citet{sheth03}. We also consider the
effect of  redshift evolution of the velocity function by using a
simple parametric model (see \S \ref{sec:gal} for details).  
\item The condition $i_{\rm qso}-i_{\rm gal}<0$ may involve some
uncertainties in estimating the galaxy luminosity or the condition
itself. Thus we allow an uncertainty of $\pm0.5$ in the cut to
estimate the systematic error that this introduces. The lens sample is
unaffected by this change.
\end{itemize}

\begin{deluxetable}{ccccc}
\tablecaption{Systematic errors\label{tab:syst}}
\tablewidth{0pt}
\tablehead{
 \colhead{}  & \colhead{$w=-1$} &
 \colhead{\hspace*{1mm}} & 
 \multicolumn{2}{c}{$w\neq -1$ (with BAO)}\\
 \cline{2-2} \cline{4-5} 
\colhead{Uncertainty} & \colhead{$\Delta\Omega_\Lambda$}  
& \colhead{} & \colhead{$\Delta\Omega_{\rm M}$} & \colhead{$\Delta w$} }
\startdata
prolate $25\%-75\%$  & $\pm0.04$ & & $\pm0.02$ & $^{+0.1}_{-0.2}$ \\
$\bar{e}\rightarrow \pm 0.1$ & $-0.01$ & &  $+0.00$ & $+0.0$ \\
$\beta_l\rightarrow \pm 0.2$ & $\pm0.04$ & & $\pm0.02$ & $\pm0.2$ \\
external shear & $+0.00$ & & $-0.00$ & $-0.0$ \\
external convergence & $-0.01$ & & $+0.01$ & $+0.1$ \\
Sheth et al. $dn/d\sigma_v$ & $+0.10$ & & $-0.04$ & $-0.4$ \\
$dn/d\sigma_v$ evolution & $+0.04$ & & $-0.02$ & $-0.2$ \\ 
$i_{\rm qso}-i_{\rm gal}\rightarrow \pm 0.5$ 
& $^{+0.04}_{-0.02}$ & & $\pm0.01$ & $\pm 0.1$ \\ \hline
total & ${}^{+0.13}_{-0.06}$ & & ${}^{+0.03}_{-0.05}$ & ${}^{+0.3}_{-0.5}$ \\
\enddata
\tablecomments{Total errors are estimated from the quadrature sum of all errors.}    
\end{deluxetable}

Our results summarized in Table \ref{tab:syst} indicate that the
largest uncertainties in our conclusions come from the dynamical
normalization, the faint end of the quasar luminosity functions, 
the velocity function of the lens galaxies and its redshift
evolution. In particular, changing the velocity function significantly
increases the best-fit value of the cosmological constant, as
discussed by \citet{chae07}. In essence, the lensing optical depth
scales as $\sigma_*^4$, so small errors in the velocity function lead
to much larger errors in the cosmological estimates.  Ideally we would
self-calibrate the velocity function based on the image separations
\citep[see][]{kochanek93}, but the two models we consider here have
very similar predictions for the image separations (see Figure
\ref{fig:sep}) and the data only favor the \citet{choi07} model by the
$\chi^2$ difference of $\sim 0.5$. Our result indicates that the
errors from the galaxy ellipticity and external perturbations are
negligibly small.  If we combine all these uncertainties, we obtain
systematic errors comparable to the statistical errors, e.g.,
$\Omega_\Lambda=0.74^{+0.11}_{-0.15}({\rm stat.})^{+0.13}_{-0.06}({\rm
  syst.})$ for the cosmological constant case. 

\section{Summary and discussion}\label{sec:summary}

We have derived constraints on dark energy using the new optical
strong lens sample from the SQLS DR3. We take various
selection effects into account to make reasonably robust predictions
for the number of lensed quasars in the SQLS. We have found that the
derived constraints agree well with the current concordance
cosmology. Assuming a cosmological constant ($w=-1$) we have obtained
$\Omega_\Lambda=0.74^{+0.11}_{-0.15}({\rm stat.})^{+0.13}_{-0.06}({\rm
  syst.})$. The constraint primarily arises from the total number of
lenses in our statistical sample. For the more general case $w\neq-1$,
the constraint was combined with that of the SDSS BAO
\citep{eisenstein05} to break the degeneracy. The resulting joint
constraint is $\Omega_{\rm M}=0.26^{+0.07}_{-0.06}({\rm
  stat.})^{+0.03}_{-0.05}({\rm syst.})$ and $w=-1.1\pm0.6({\rm
  stat.})^{+0.3}_{-0.5}({\rm syst.})$. The results are in good
agreement with recent constraints from radio lenses
\citep{chae03a,chae07,mitchell05}. The results confirm the current
standard picture that the universe is dominated by dark energy with a
cosmological constant-like equation of state, independently of type Ia
supernovae.   

Although we have quantified our systematics on our cosmological
results by changing several important assumptions, there are
additional systematic effects that could change our quantitative
results. One is dust extinction by lensing  galaxies. Previous studies
have shown that dust is indeed present in lensing galaxies even if
they are early-type galaxies, although the measured total extinction
is modest, $E(B-V)\sim 0.1$~mag 
\citep[e.g.,][]{falco99}.  Since we set the magnitude limit at the
$i$-band for which dust extinction is less important than in the bluer
bands, we expect that this effect will not heavily influence our
result. Moreover, the flux from the lensing galaxy 
slightly increases the PSF magnitude of the lens system, and this effect
could compensate the effect of dust extinction to some extent. 
Another effect we have neglected is lensing by multiple galaxies.
Theoretically the probability for such multiple lensing is just a few
percent of the lensing probability by a single galaxy at $z\lesssim2$
\citep{moeller01} and therefore can be neglected. However, it only
considered chance superpositions along the line of sight and ignored
nearby correlated galaxies that could dominate the contribution to
multiple lens events \citep[e.g.,][]{cohn04}. Moreover, the fact that
one of our lens sample, SDSS J1001+5027 \citep{oguri05a}, has a
secondary galaxy near the primary lensing galaxy suggests that the
effect needs to be addressed carefully.   

In this paper we have assumed a flat universe. Although this is a
reasonable assumption given that virtually all current cosmological
constraints are consistent with a flat universe
\citep[e.g.,][]{tegmark06,ichikawa07,spergel07,wang07,allen07}, it is
still of interest to consider non-flat universes. In particular the
introduction of both non-flatness and $w\neq -1$ results in additional
strong degeneracy between cosmological parameters
\citep[e.g.,][]{linder05}, and thus may require an additional
independent cosmological probe in order to obtain tight constraints on
individual parameters.  

Cosmological constraints presented in this paper are obtained from the
SQLS DR3 sample, which represents $\sim 40\%$ of the full SDSS
data. The extension of the SQLS lens sample to the SDSS DR5 and
SDSS-II is in progress, therefore we expect that the statistical
errors will improve significantly in the near future. Better
constraints may be obtained by considering the SQLS and CLASS samples
jointly. In addition to the extension of the lens sample, it is of
great importance to reduce the systematic errors.  An important
advantage of our optical survey over radio surveys is that there is no
systematics from the source redshift distribution, which was the
biggest source of systematic error in the CLASS analysis
\citep{chae03a}. The velocity function of galaxies and the faint end
quasar luminosity function are expected to converge in the near future
as current large-scale surveys are completed, thus we expect the
systematic errors can be reduced in future analyses of the SQLS
lenses. If we have a large enough sample of lenses, we may be able to
reduce the systematic effect further by calibrating the velocity
function from the observed image separation distribution itself.   

Although in this paper we have restricted ourselves to galaxy-scale
lenses to study dark energy, our statistical lens sample contains
group- or cluster-scale lenses as well. The number of cluster-scale
lenses in our sample is quite sensitive to the abundance of clusters
at intermediate redshifts ($z\sim 0.5$), and therefore was used to
study $\sigma_8$ \citep{oguri04,li07}. The full image separation
distribution from galaxy- to cluster-scales will be valuable in
understanding how galaxies are populated in dark matter halos. 

\acknowledgments

We thank Kyu-Hyun Chae for useful discussions.  
This work was supported in part by the Department of Energy contract
DE-AC02-76SF00515. 
N.~I. acknowledges supports from the Japan Society for the Promotion of 
Science and the Special Postdoctoral Researcher Program of RIKEN. 
M.~A.~S. acknowledges support from NSF grant AST 03-07409.
A portion of this work was also performed under the auspices of the U.S. 
Department of Energy, National Nuclear Security Administration by the
University  of California, Lawrence Livermore National Laboratory under
contract No. W-7405-Eng-48. 
I.~K. acknowledges supports from Ministry of Education, Culture, Sports, 
Science, and Technology, Grant-in-Aid for Encouragement of Young
Scientists (No. 17740139), and Grant-in-Aid for Scientific Research on
Priority Areas No. 467 ``Probing the Dark Energy through an Extremely
Wide \& Deep Survey with Subaru Telescope''.
A.~C. acknowledges the support of CONICYT, Chile, under grant FONDECYT
1051061.

Funding for the SDSS and SDSS-II has been provided by the Alfred
P. Sloan Foundation, the Participating Institutions, the National
Science Foundation, the U.S. Department of Energy, the National
Aeronautics and Space Administration, the Japanese Monbukagakusho, the
Max Planck Society, and the Higher Education  Funding Council for
England. The SDSS Web Site is http://www.sdss.org/. 

The SDSS is managed by the Astrophysical Research Consortium for the
Participating Institutions. The Participating Institutions are the
American Museum of Natural History, Astrophysical Institute Potsdam,
University of Basel, Cambridge University, Case Western Reserve
University, University of Chicago, Drexel University, Fermilab, the
Institute for Advanced Study, the Japan Participation Group, Johns
Hopkins University, the Joint Institute for Nuclear Astrophysics, the
Kavli Institute for Particle Astrophysics and Cosmology, the Korean
Scientist Group, the Chinese Academy of Sciences (LAMOST), Los Alamos
National Laboratory, the Max-Planck-Institute for Astronomy (MPIA),
the Max-Planck-Institute for Astrophysics (MPA), New Mexico State
University, Ohio State University, University of Pittsburgh,
University of Portsmouth, Princeton University, the United States
Naval Observatory, and the University of Washington.

\end{document}